\RequirePackage{fix-cm}
\documentclass[twocolumn]{svjour3}          
\smartqed  
\usepackage{graphicx}
\usepackage{caption}
\usepackage{color}
\usepackage{soul}
\usepackage{flushend}
\usepackage{amsmath}
\usepackage{amssymb,amsfonts}

\usepackage{multirow}
\usepackage{framed} 
\usepackage{multicol} 
\usepackage{nomencl} 
\usepackage{comment}
\usepackage{algorithm} 
\usepackage{algpseudocode}

\newcommand*{\affmark}[1][*]{\textsuperscript{#1}}
\begin{document}

\title{A Modified Epidemiological Model to Understand the Uneven Impact of COVID-19 on Vulnerable Individuals and the Approaches Required to Help them Emerge from Lockdown
}

\titlerunning{A Modified Epidemiological Model to Understand the Impact of COVID-19 on Vulnerable Individuals}        

\author{Dario Ortega Anderez\protect\affmark[*1] \and Eiman Kanjo\affmark[1] \and Ganna Pogrebna\affmark[3] \and Shane Johnson\affmark[2] \and John Alan Hunt\affmark[4] 
}


\institute{Dario Ortega Anderez and  Eiman Kanjo\protect\affmark[1]School of Science and Technology, Nottingham Trent University, Clifton Lane, NG11 8NS, UK.
            \email{dario.ortegaanderez02@ntu.ac.uk}
            \email{eiman.kanjo@ntu.ac.uk}
\and 
    Shane Johnson \at
              \protect\affmark[2]University College London (UCL) Jill Dando Institute, 
35 Tavistock Square, 
London, 
WC1H 9EZ, UK 
\and 
           Ganna Pogrebna\at
              \protect\affmark[3]The University of Sydney Business School, Abercrombie Building H70, Darlington NSW 2006, Australia and the Alan Turing Institute, 96 Euston Road, London, NW1 2DB, UK
\and 
    John Hunt \at
              \protect\affmark[4]Medical  Technologies  Innovation  Facility,  Nottingham  Trent  University, Nottingham,  NG11  8NS,  U.K. 
                College  of  Biomedical  Engineering,  China  Medical  University,  Taichung 40402,  Taiwan
}

\date{Received: date / Accepted: date}
\maketitle

\begin{abstract} 

\hfill
\hfill

\noindent \textbf{Background:} COVID-19 has shown a relatively low case fatality rate in young healthy individuals, with the majority of this group being asymptomatic or having mild symptoms. However, the severity of the disease among the elderly as well as in individuals with underlying health conditions has caused significant mortality rates worldwide. Understanding this variance amongst different sectors of society and modelling this will enable the different levels of risk to be determined to enable strategies to be applied to different groups. Long-established compartmental epidemiological models like SIR and SEIR do not account for the variability encountered in the severity of the SARS-CoV-2 disease across different population groups. 

\noindent \textbf{Objective:} The objective of this study is to investigate how a reduction in the exposure of vulnerable individuals to COVID-19 can minimise the number of deaths caused by the disease, using the UK as a case study.    

\noindent \textbf{Methods:} To overcome the limitation of long-established compartmental epidemiological models, it is proposed that a modified model, namely SEIR-v, through which the population is separated into two groups regarding their vulnerability to SARS-CoV-2 is applied. This enables the analysis of the spread of the epidemic when different contention measures are applied to different groups in society regarding their vulnerability to the disease. A Monte Carlo simulation (100,000 runs) along the proposed SEIR-v model is used to study the number of deaths which could be avoided as a function of the decrease in the exposure of vulnerable individuals to the disease. 

\noindent \textbf{Results:} The results indicate a large number of deaths could be avoided by a slight realistic decrease in the exposure of vulnerable groups to the disease. The mean values across the simulations indicate 3,681 and 7,460 lives could be saved when such exposure is reduced by 10\% and 20\% respectively. 

\noindent \textbf{Conclusions:} From the encouraging results of the modelling a number of mechanisms are proposed to limit the exposure of vulnerable individuals to the disease. One option could be the provision of a wristband to vulnerable people and those without a smartphone and contact-tracing app, filling the gap created by systems relying on smartphone apps only. By combining very dense contact tracing data from smartphone apps and wristband signals with information about infection status and symptoms, vulnerable people can be protected and kept safer. 
\end{abstract}

\section{Introduction}
\label{intro}
Coronaviruses (CoV) are a large family of enveloped, positive-strand RNA viral diseases capable of infecting a variety of host species, including humans and several other vertebrates \cite{Channappanavar2017}. CoVs predominantly cause gastrointestinal and respiratory tract infections, inducing a wide range of clinical symptotic manifestations \cite{Su2016}.

Before the latest strain of coronavirus SARS-CoV-2 in December 2019, six human CoVs including four endemic (HCoV-OC43, -229E, -NL63, and -HKU1) and two epidemic (SARS-CoV and MERS-CoV) viruses had been identified. 

Despite the recent scientific progress across different domains and the increasing levels of public hygiene worldwide, the combination of various key adverse factors has translated the SARS-CoV-2 outbreak into a global epidemic with a real risk to life. The COVID-19 outbreak originated in Wuhan, China, which in addition to being the largest city in central China, is home to the largest deep-water port, airport and train station in that same area \cite{Wilson2020}. The outbreak coincided with increased levels of travel prior to the Chinese New Year, that resulted in more than five million people travelling out from Wuhan in the first few days of the outbreak before Wuhan was put under lockdown \cite{Wilder2020}. This facilitated the spread of COVID-19 within Mainland China as well as in countries with high volumes of air traffic with China, such as South Korea, Japan, Thailand and Singapore \cite{Bogoch2020}. Initial estimates from the World Health Organisation (WHO) for the basic reproduction number ($R_0$) of COVID-19 were in the range of 1.4 to 2.5, whereas recent studies suggest this estimated basic reproduction number could have been higher, with statistical studies giving $R_0$ values in the range (2.2-3.58) \cite{Liu2020,Zhao2020,Li2020}, mathematical studies from 1.4 to 6.49 \cite{Shen2020,Cao2020,Tang2020}, and stochastic approaches from 2.2 to 2.68 \cite{Wu2020,Riou2020}. This implies the reproduction number of COVID-19 may be a lot higher than originally estimated and therefore significantly greater than that of MERS ($R_0<1$) \cite{World2017} and in the range of SARS' ($R_0\approx3$) \cite{World2003}. As the COVID-19 epidemic unfolds, the commonality of mild and asymptomatic cases has become apparent. This is a key issue, since unlike with other CoVs like SARS, where peak viral shedding occurred after patients were severely ill and therefore easily identifiable, preliminary evidence suggests that COVID-19 can be transmitted at the early phases of the disease where no symptoms are observed \cite{Rothe2020}.     

The combination of these factors has led to the rapid international spread of SARS-CoV-2 and it being declared a Public Health Emergency of International Concern on 1 February 2020 and thereafter a pandemic on the 11 March 2020. The dramatic consequences of the pandemic have translated into unprecedented non-pharmaceutical interventions (NPIs) internationally, including national and international travel restrictions, quarantine and isolation of large populations, the closure of universities, schools and social spaces, as well as a widespread social distancing policy and the banning of public gatherings and events. Despite these containment measures by different countries, as of 2 June 2020, over 6.4 million cases of COVID-19 have been reported worldwide causing death to over 378,290 people \cite{WorldometersWorld}.

While COVID-19 has the potential to infect every individual in the world, the disease has been extremely dangerous with a greater risk of mortality in groups older than 70 years of age as well as in people with underlying health conditions (e.g., high blood pressure; respiratory problems, etc.). Given the current increasingly aging population in many parts of the world, particularly, in developed nations such as the UK, where 2018 figures from the Office for National Statistics \cite{Coates2019} reported that 18.3\% of people were 65 or over. Put this, with the number of younger individuals who have underlying health conditions, then there is an approximate 20\% of the UK population having a considerably high risk of death if infected.    
Motivated by the disparate mortality rates across different groups in society, this work proposes a modified compartmental epidemiological model (SEIR-v) to study the impact of reducing the contact rate of vulnerable individuals on the potential to reduce the number of fatalities caused by the disease. The motivation behind the model was that traditional epidemiological models like SIR \cite{Kermack1927} and SEIR \cite{Aron1984} assume equal contact rates and death rates for every individual in the population (i.e., these models, essentially, abstract from significant individual, behavioral, and spacial heterogeneity observed in the population of any country). In contrast, SEIR-v provides a means of studying the progression of the number of fatalities when differing contention measures are applied across different groups of individuals regarding their vulnerability to the disease, while also accounting for the characteristic variability seen in the case fatality rates across these groups. As outlined in \cite{Ferretti2020}, the original mathematical models the British government used to inform policy in the UK did not account for vulnerable people. In contrast, SEIR-v provides the opportunity to distinguish between vulnerable groups and low-risk groups, allowing policy recommendations for each segment, rather than applying a one-size-fits-all policy. 
Predictions made with SEIR-v outlined the importance of minimising the chances of vulnerable people contracting the disease, with an estimated reduction of 3681 and 7406 further deaths if their exposure to the virus was decreased by only 10\% and 20\% respectively. 
The widespread policies of physical distancing restrictions, may now have contributed to another health problem: loneliness. Those who do not have close family or friends, and rely on the support of voluntary services or social care, may feel vulnerable, according to recent correspondence published in The Lancet \cite{Armitage2020}. The impact on physical health, arising from mental health problems due to loneliness should not be underestimated and ignored.

There is therefore a need to help vulnerable people exit the lockdown whilst addressing their continued protection and provide them with the means to participate in the contact-tracing process.
Here, we consider practical options for facilitating the exit strategy for vulnerable groups from lockdown. In line with this, we provide a set of recommendations:
\begin{enumerate}
  \item the use of wearable devices (henceforth, wearables) to enable vulnerable people to take part in contact tracing, 
  \item the development of effective incentive mechanisms in order to motivate people to engage in contact tracing,
  \item the use of digital tools to maintain physical distancing and monitor health symptoms, 
  \item the use of personal protective equipment, 
  \item the planning for easing the contention measures.  
\end{enumerate}
The most important determinants of outcome are:
\begin{enumerate} 
\item reduction of transmission rates post lockdown in the vulnerable populations; 
\item fewer restrictions on the vulnerable post-lockdown with noticeable improvement in their well-being (many may already be suffering from loneliness and mental health problems due to the lockdown);
\item maintenance of keeping vulnerable people and the hard-to-reach connected and closely monitored.
\end{enumerate}

\section{Methods}
\subsection{Compartmental Epidemiological Models}\label{sec:Models}
\begin{figure}[t]
  \centering
  \includegraphics[width=0.48\textwidth]{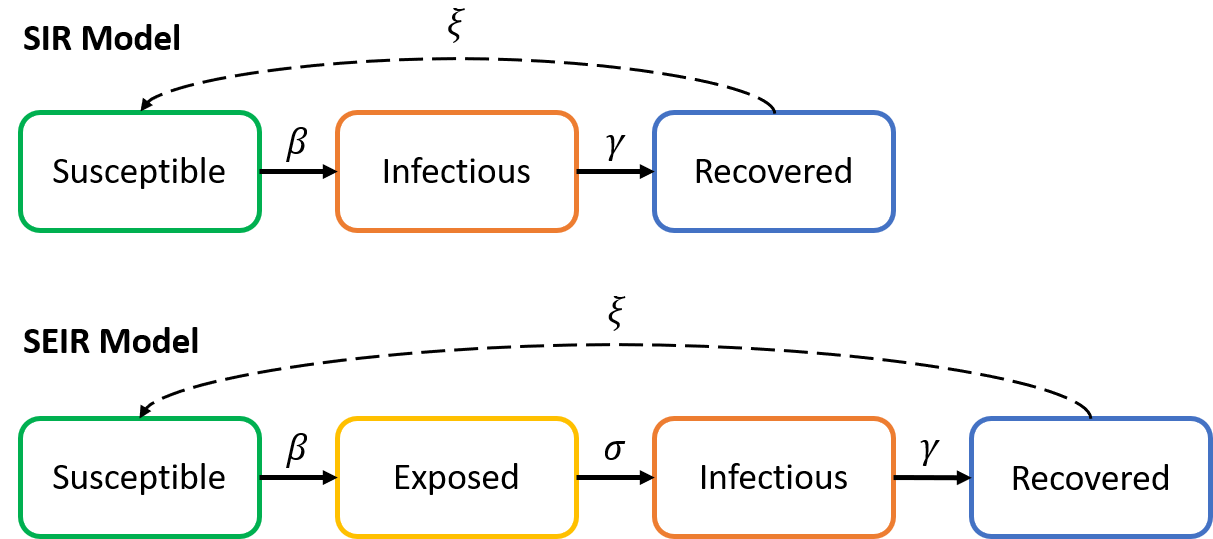}
  \caption{SIR and SEIR epidemiological models.}
  \label{fig:SIRSEIR}
\end{figure}

Generally, the spread of infectious diseases like COVID-19 are studied through the use of compartmental epidemiological models. Such models provide a simplified means of describing the transmission of the infectious disease through the different individuals within a population by dividing the individuals into different states regarding their current susceptibility to the disease and their disease transmission capabilities. Within such models, the SIR \cite{Kermack1927} and the SEIR \cite{Aron1984} models are widely employed and published in the literature in the field. A visual representation of the SIR and SEIR models is depicted in Figure \ref{fig:SIRSEIR}. 

\subsubsection{SIR Model}\label{SIR}
The SIR model divides the entire population into three different compartments, namely susceptible, infectious and recovered. The transition between one compartment to its following are controlled by the model transition parameters:
\begin{enumerate}
  \item Infectious rate ($\beta$): is the rate of spread of the virus given by the probability of transmitting the disease between an infectious individual and a susceptible individual. This is subject to the disease transmission probability and the chance of contact.
  \item Recovery rate ($\gamma$) = $\frac{1}{T_{lat}}$ is determined by the average duration of the infectious period of the disease ($T_{lat}$). 
  \item Re-susceptibility rate ($\xi$) is the rate at which recovered individuals return to the susceptible state due to loss of immunity (normally ignored due to long-term immunity).
\end{enumerate}

Given the definition of the above parameters, the SIR model can be expressed by:
\begin{equation}
  \begin{aligned}
  \frac{\partial S}{\partial t}=-\frac{\beta IS}{N},\\\\ 
  \frac{\partial I}{\partial t}=\frac{\beta IS}{N}-\gamma I,\\\\
  \frac{\partial R}{\partial t}=\gamma I
  \end{aligned}
\end{equation}

\noindent where $S$, $I$ and $R$ are the number of susceptible, infected and recovered (or removed) individuals respectively, and N is the total population, which follows $N=S(t)+I(t)+R(t)$.

Provided the set of equations above, the dynamics of the infectious disease given by its reproduction number ($R_
0$) is calculated as:
\begin{equation}
  R_0=\frac{\beta}{\gamma}  
\end{equation}

\subsubsection{SEIR Model}\label{SEIR}
The SEIR model proposed in \cite{Aron1984} is a slight variation of the SIR model, in that it includes an additional state `Exposed' to the three states used in the SIR model. The motivation behind this comes from the fact that some infectious diseases exhibit a considerable post-infection incubation period in which an infected person (exposed) is not yet infectious, thereby affecting significantly the dynamics of the transmission of the disease. The SEIR model can be expressed as follows:

\begin{equation}
\begin{aligned}
  \frac{\partial S}{\partial t}=-\frac{\beta IS}{N},\\\\ 
  \frac{\partial E}{\partial t}=\frac{\beta IS}{N}-\sigma E ,\\\\
  \frac{\partial I}{\partial t}=\sigma E - \gamma I,\\\\
  \frac{\partial R}{\partial t}= \gamma I
\end{aligned}
\end{equation}

\subsection{The Role of Non Pharmaceutical Interventions and Herd Immunity}\label{sec:interventions}

Non Pharmaceutical Interventions (NPIs) are those actions or measures employed with the aim of limiting the spread of a viral disease when pharmaceutical interventions, such as anti-viral medications and vaccines, are still not available. A general classification into two main contention strategies, namely suppression and mitigation, is made based on whether the measures applied aim at quickly reducing the reproduction number (the average number of secondary cases generated per typical infectious case), $R$, to values lower than 1, or to simply slow down the spread of the virus by controlling the value of $R$, while allowing it to take values in ($R\geq1$).

In other words, suppression strategies aim to turn the pandemic phase of the disease ($R>1$) into the endemic phase ($R<1$), where each infectious individual, on average, spreads the virus to less than one person, thereby causing a decay in the daily number of new cases. In contrast, mitigation strategies, unless combined with certain levels of population immunity, are not aimed at the suppression of the virus \textit{per se}. Instead, they are employed to reduce the health impact of the epidemic by controlling the curve through the contention of $R$, so that even though each individual, on average, spreads the disease to more than one person, such spread is to some extent controlled to meet the capacity of the respective health care system, while building up population immunity along the course of the epidemic phase. Ultimately, such built up immunity will prevent the disease from spreading any further, leading thereby to a rapid decline in the number of new infections and the consequent endemic phase of the disease. 

\begin{figure}[t]
  \centering
  \includegraphics[width=0.48\textwidth,height=0.12\textheight]{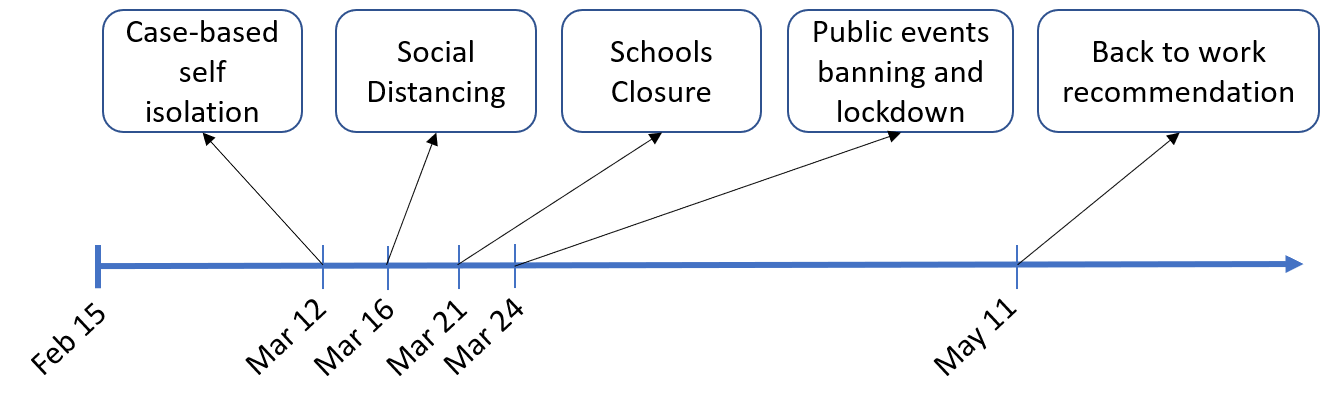}
  \caption{Non-pharmaceutical interventions applied by the UK government.}
  \label{fig:wristband}
\end{figure}

To date, hybrid strategies through which mitigation and suppression measures are combined to fight the spread of the epidemic, are being adopted by the vast majority of countries severely affected by COVID-19. Such strategies have overall included the implementation of diverse contention measures including the instruction to self-isolate to confirmed and suspected cases, the encouragement of social distancing, the banning of non essential travel, mass gatherings and public events, the closure of schools and universities and the lockdown of the population. Whilst allowing key workers to carry out their duties and then broadly the general public in lockdown at home were allowed to go out for essentials like food and exercise.     
The analysis of existing data has already demonstrated that mass testing and the isolation of infected individuals can on its own have a suppressive impact on the curve, thereby reducing significantly the size of the peak \cite{Aleta2020}. Examples include the strategies followed by South Korea and Germany. However, it must be noted that the adoption of suppression strategies is challenging due to the low level of herd immunity achieved as function of time throughout the progression of the disease, thereby, to avoid the eventual increase of new infections, the contention measures have to be maintained until pharmaceutical interventions are available.   

\begin{figure*}[t]
  \centering
  \includegraphics[width=0.65\textwidth]{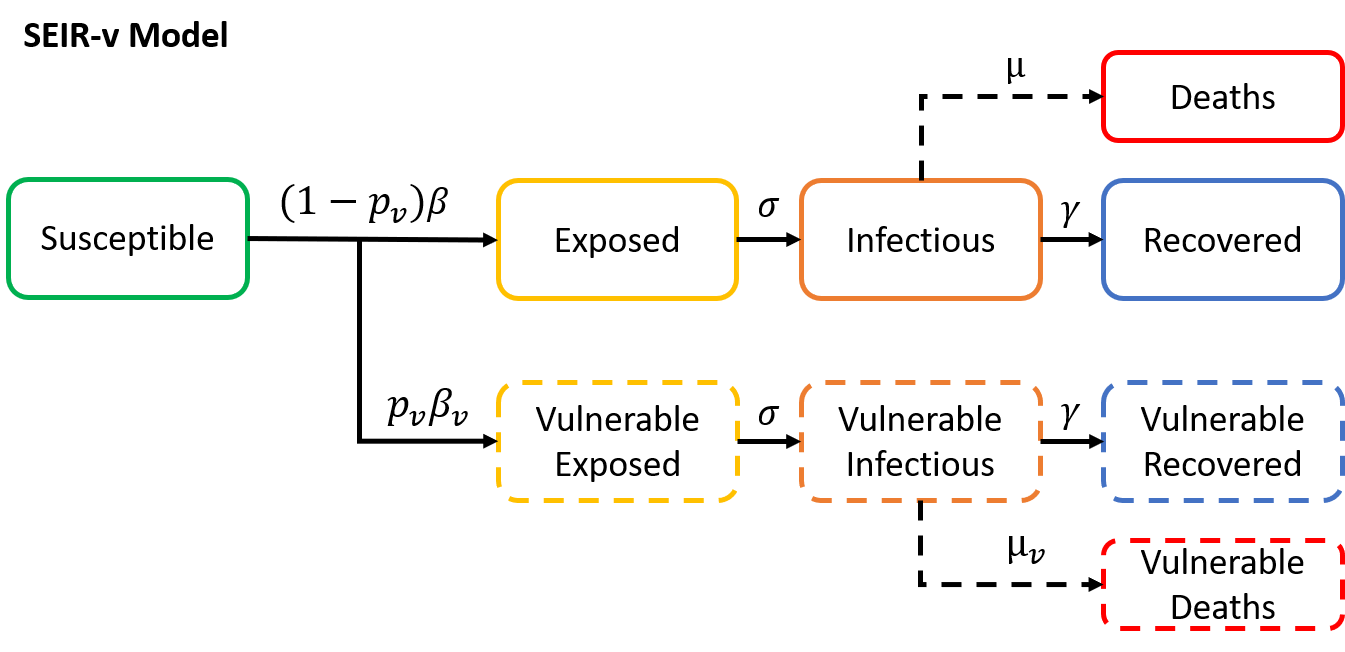}
  \caption{SEIR-v epidemiological model.}
  \label{fig:SEIR-v}
\end{figure*}

Although still no empirical evidence can be found suggesting the dismissal of the possibility of reinfection from COVID-19, the results reported on studies investigating the persistence of antibodies in patients exposed to similar CoVs \cite{Mo2006,Wu2007,Alshukairi2016,Payne2016} suggest the antibody immunity built up by individuals exposed to the SARS-CoV-2 virus may potentially last until medical interventions are available. For instance, the study in \cite{Mo2006} showed the persistence of antibodies in patients infected by the severe acute respiratory syndrome coronavirus (SARS-CoV), could last for at least two years. The results reported in \cite{Wu2007}, where 176 patients were found to maintain SARS-specific antibodies for 2 years were in line with those of \cite{Mo2006}. A significant reduction of immunoglobulin G–positive was the case in the third year. Thus, SARS patients might be susceptible to reinfection 3 years after the initial exposure to the virus. Regarding the immunity against MERS-CoV, the work in \cite{Alshukairi2016} studied the antibody response in 9 healthcare workers in Jeddah, Saudi Arabia, who had previously suffered from the disease showing symptoms of severe pneumonia. A further study in \cite{Payne2016}, explored the long-term antibody response against MERS-CoV. This research reported the persistence of antibodies, including neutralising antibodies, in 6 out of the 7 (86\%) explored individuals for at least 34 months after the outbreak.

\subsection{Proposed Model: SEIR-v}\label{sec:proposed_model}
The need to develop the SEIR-v model comes from the consideration of the characteristic variability encountered in the severity of COVID-19 across different individuals with respect to their age group and state of health during the early phase of the infection. Evidence suggests, the mortality rate in individuals of advanced age and / or with underlying health conditions is significantly higher than that in younger healthier individuals \cite{Ferguson2020}. Thereby, to reduce the number of fatalities, it is crucial to have detailed means of looking more closely into the impact of the application of the different NPIs across the different population groups regarding their vulnerability to the disease on the overall mortality rate. SEIR-v provides a means of studying the impact of the different NPIs on the number of deaths, when these are only applied to people with higher vulnerability to the disease. The SEIR-v compartmental model can be defined as follows:

\begin{equation}
\begin{split}
  \frac{\partial S}{\partial t}=-(1-p_v)\frac{\beta IS}{N} - p_v\frac{\beta_v IS}{N},\\\\ 
  \frac{\partial E}{\partial t}=(1-p_v)\frac{\beta IS}{N}-\sigma E\\\\
  \frac{\partial E_v}{\partial t}=p_v\frac{\beta_v IS}{N}-\sigma E_v,\\\\
  \frac{\partial I}{\partial t}=\sigma E - \gamma I,\\\\
  \frac{\partial I_v}{\partial t}=\sigma E_v - \gamma I_v,\\\\
  \frac{\partial R}{\partial t}= \gamma I(1-\mu),\\\\
  \frac{\partial R_v}{\partial t}= \gamma I(1-\mu_v),\\\\
  \frac{\partial D}{\partial t}= \gamma I\mu + \gamma I_v\mu_v
\end{split}
\end{equation}

\noindent where $S$ is the number of susceptible individuals. $E$, $I$, $R$ are the number of non vulnerable exposed, infectious and recovered individuals respectively. $E_v$ $I_v$, $R_v$ are the number of vulnerable exposed, infectious and recovered individuals respectively. $D$ is the total number of deaths caused by the disease. $p_v$ is the probability of an individual being vulnerable to the disease. $\beta$ and $\beta_v$ are the contact rates by non vulnerable and by vulnerable individuals respectively. $\sigma$ is the rate at which an exposed individual becomes infectious. $\gamma$ is the the rate at which an infectious individual recovers from the disease. $\mu$ and $\mu_v$ are the case fatality rates for non vulnerable and for vulnerable individuals respectively. The conceptual diagram of SEIR-v can be seen in Figure \ref{fig:SEIR-v}.

\begin{table*}[t]
\caption{Description of the SEIR-v model parameters}
\centering
\renewcommand{\arraystretch}{2.0}
\resizebox{\textwidth}{!}{\begin{tabular}{|l|l|l|l|}

\hline
\textbf{Parameter} & \textbf{Description}                                                              & \textbf{Value}                                              & \textbf{Comments}                                                                           \\ \hline
N         & Population                                                               & 67.838.235 \cite{WorldometersUK}  & Total population in the UK as of 2020                                              \\ \hline

$E_{v_0}$   & Vulnerable Exposed                                                       & 2                                                   & Vulnerable individuals exposed to the disease at the beginning of the outbreak     \\ \hline
$E_0$     & Exposed                                                                  & 4                                                 & Non-vulnerable individuals exposed to the disease at the beginning of the outbreak \\ \hline
$I_{v_0}$     & Infected                                                                  & 0                                                  & Vulnerable infected individuals at the beginning of the outbreak \\ \hline
$I_0$     & Infected                                                                  & 1                                                 & Non-vulnerable infected individuals at the beginning of the outbreak \\ \hline
$T_{inc}$ & Incubation period                                                        & 5.6 \cite{Linton2020}             & $\sigma = \frac{1}{T_{inc}}$ The time it takes for an exposed individual to become infectious                                                      \\ \hline
$T_{lat}$ & Latent period                                                            & 7.5 \cite{Linton2020}             & $\gamma = \frac{1}{T_{lat}}$ The time it takes for an infectious individual to recover                                                             \\ \hline
$\mu_v$   & \begin{tabular}[c]{@{}l@{}}Vulnerable\\ Case Fatality Rate \end{tabular}      & [0.005-0.037, 95\% CI]\%  & Case fatality rate of COVID-19 on vulnerable individuals                               \\ \hline
$\mu$     & \begin{tabular}[c]{@{}l@{}}Non-vulnerable \\ Case Fatality Rate\end{tabular} & [0.000007-0.000011, 95\% CI]\%    & Case fatality rate of COVID-19 on non-vulnerable individuals                           \\ \hline
$p_v$   & Vulnerable probability & 0.2  & Probability of an individual being vulnerable to the disease \\ \hline
$\eta$   & Fear Factor & 0.33 \    & \begin{tabular}[c]{@{}l@{}@{}} Fear factor caused by the recommendation made by the UK government for \\ vulnerable individuals to stay at home for at least 12 weeks at the beginning \\of the outbreak and the widespread severity of the disease within this group \end{tabular} \\ \hline
$\beta_{0}$ & Initial Contact Rate                                                            & [0.5-2.1, 95\% CI]              & Contact rate at the beginning of the outbreak                                                             \\ \hline
$\beta_{1}$ & Contact Rate 1                                                           & [0.9-0.95, 95\% CI]*$\beta_0$              & Contact rate after the mandate of case-based self isolation                                                               \\ \hline
$\beta_{2}$ & Contact Rate 2                                                           & [0.9-0.95, 95\% CI]*$\beta_1$             & Contact rate after government encouragement for social distancing                                                               \\ \hline
$\beta_{3}$ & Contact Rate 3                                                            & [0.75-0.85, 95\% CI]*$\beta_2$              & Contact rate after schools closure                                                              \\ \hline
$\beta_{4}$ & Contact Rate 4                                                           & [0.40-0.60, 95\% CI]*$\beta_3$             & Contact rate after lockdown order and banning of public events                                                             \\ \hline
$\beta_{5}$ & Contact Rate 5                                                           & [1.1-1.9, 95\% CI]*$\beta_4$             & Contact rate after recommendation for people to go back to work                                                              \\
\hline
$\beta_{v_i}$ & Vulnerable Contact Rate                                                           & $\eta*\beta_i$             & Contact rate of vulnerable individuals.                                                              \\
\hline

\end{tabular}}
\label{tab:parameters}
\end{table*}

\subsection{Model Parameterisation}\label{parameterisation}
Definitive values for the different parameters that define the spread of the COVID-19 and the impact of the contention measures in place are still unknown, given that the disease is still spreading globally and key information is still unknown. For instance, the number of asymptomatic cases and consequently the total number of individuals who are or have been infected by the disease remains to be determined. Consequently, the case fatality rate of the disease is also unknown. Motivated by this fact, first efforts were aimed at estimating the value of the different model parameters through the comparison of the estimated number of deaths and the real number of deaths as reported in \cite{WorldometersUKDeaths}.To do so, the incubation period ($\sigma$) and the recovery rate ($\gamma$) estimated in previous work were taken into consideration \cite{Linton2020}. With these figures, the best fit parameters were identified by running a Monte Carlo simulation with 100.000 iterations, allowing the rest of the parameters to take random values from a Gaussian distribution within their respective expected intervals. The definitions of the different model parameters can be found in Table \ref{tab:parameters}.

\subsubsection{Virus Transmissibility Study}
Given the parameters provided in Table \ref{tab:parameters}, the final mortality rate of COVID-19 was studied as a function of the percentage decrease in the contact rate of vulnerable individuals ($\beta_v$). The mean values across the different Monte Carlo simulations and the reduction in the number of deaths caused by the disease estimated by the best fit model are provided.

\section{Results}\label{Results}
The results achieved by the analysis of the number of deaths caused by COVID-19 as a function of the contact rate of vulnerable individuals ($\beta_v$) are presented in this section. Two different scenarios were considered:
\begin{enumerate}
    \item The vulnerable group contact rate, $\beta_v$, is decreased from the beginning of the outbreak. With this scenario, the potential reduction in the number of deaths if more protective measures for vulnerable groups had been applied from the beginning of the outbreak was studied.   
    \item The vulnerable group contact rate, $\beta_v$, is decreased from now (2 June 2020). With this, the potential reduction in the number of deaths the disease will cause from today was studied.     
\end{enumerate}

It should be noted that the Fear Factor ($\eta$) already takes into consideration the widespread risk of the disease in vulnerable groups and the recommendation made by the UK government for vulnerable individuals to stay at home for at least 12 weeks at the beginning of the outbreak. The decreases in $B_v$ are therefore additional to that caused by $\eta$. Initial simulations using higher case fatality rates significantly overestimated the number of deaths caused by the disease in the UK as compared to those reported in \cite{WorldometersUKDeaths}. It is thus believed the case fatality rate is significantly lower than that reported in \cite{Ferguson2020}.     

\begin{table*}[]
\caption{Relationship between the decrease in $\beta_v$ and the resultant number of deaths avoided when this decrease is applied from the beginning of the outbreak expressed as the mean value of the Monte Carlo simulation}
\centering
\renewcommand{\arraystretch}{1.8}
\begin{tabular}{|l|l|l|l|l|l|l|l|l|l|}
\hline
Decrease in   $\beta_v$           & 10\% & 20\%  & 30\%  & 40\%  & 50\%  & 60\%  & 70\%  & 80\%  & 90\%  \\ \hline
Decrease in number of deaths & 7699 & 15512 & 23428 & 31434 & 39519 & 47671 & 55876 & 64122 & 72395 \\ \hline
\end{tabular}
\label{tab:deathsdecrease}
\end{table*}

\subsection{Reduction of the Contact Rate of Vulnerable Individuals from the Beginning of the Outbreak}

\begin{figure}
    \centering
    \includegraphics[width=0.48\textwidth]{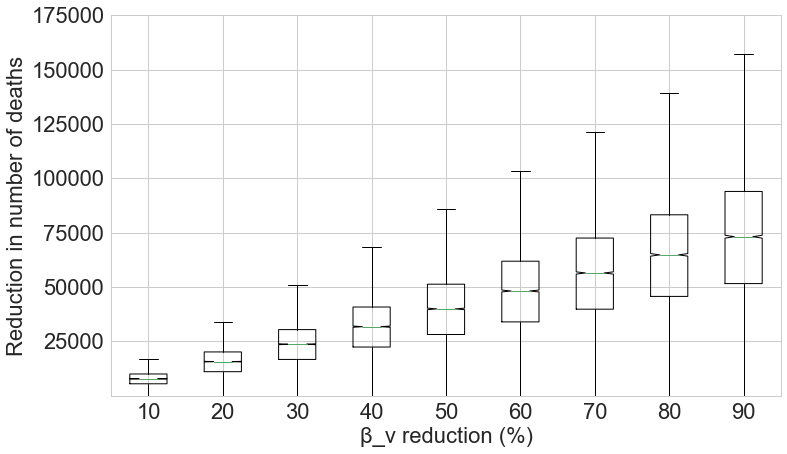}
    \caption{Reduction in the number of deaths as a function of the percentage decrease in $\beta_v$, given that this reduction is applied at the beginning of the outbreak.}
    \label{fig:betaredwhole}
\end{figure}

The potential reduction in the number of deaths caused by COVID-19 as a function of the percentage decrease applied to $\beta_v$ achieved by the Monte Carlo simulation is shown in Figure \ref{fig:betaredwhole}. From this data, the reduction of the exposure of vulnerable groups to the disease at the beginning of the outbreak greatly decreases the number of deaths, The mean figures for the prevented deaths for each 10\% decrease in $\beta_v$ are shown in Table \ref{tab:deathsdecrease}. 

\begin{figure}
    \centering
    \includegraphics[width=0.48\textwidth]{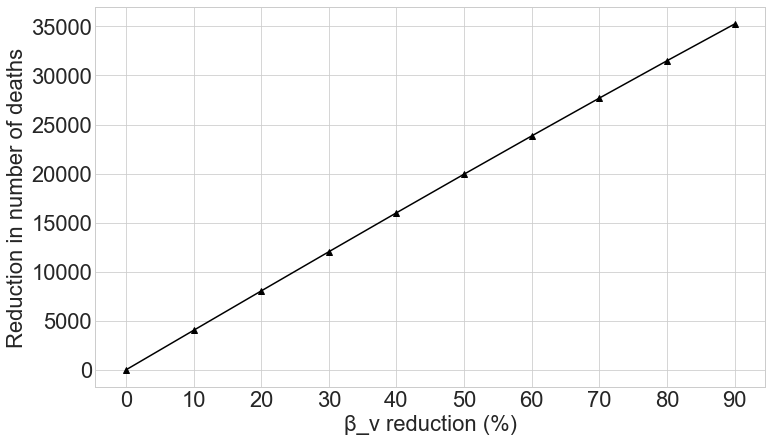}
    \caption{Reduction in the number of deaths as a function of the percentage decrease in $\beta_v$ for the best fit model, given that this reduction is applied at the beginning of the outbreak.}
    \label{fig:redbegginingmontecarlo}
\end{figure}

\begin{figure}
    \centering
    \includegraphics[width=0.48\textwidth]{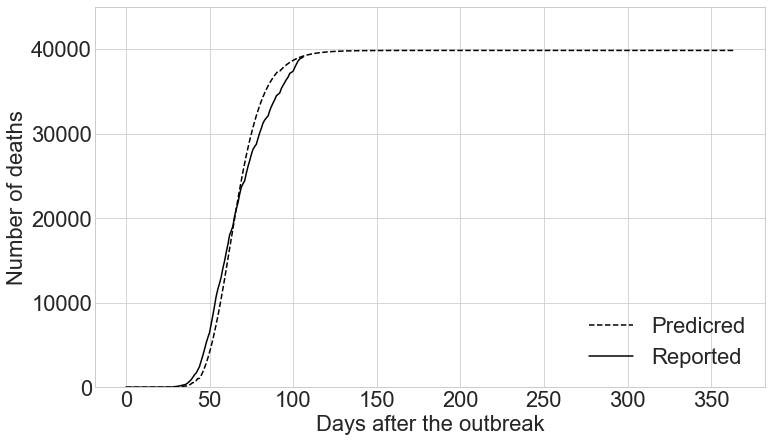}
    \caption{Predicted number of deaths using the best fit model vs the number of deaths reported (real).}
    \label{fig:bestfitmodel}
\end{figure}

Similarly, the results obtained by the use of the best fit model across the Monte Carlo simulation are shown in Figure \ref{fig:redbegginingmontecarlo}. The performance of this model at predicting the number of deaths caused by COVID-19 can be seen in Figure \ref{fig:bestfitmodel}. The prediction given by the best fit model for the total number of deaths caused by COVID-19 in the UK at the end of the outbreak is 39825.   

\subsubsection{Reduction of the Contact Rate of Vulnerable Individuals at Present}

While this data from modelling reported the decrease in the number of deaths if the exposure of vulnerable people to the virus had been reduced from the beginning of the SARS-CoV-2 outbreak, this section presents the benefits this reduction would bring if actions were taken from this point in time going forwards.  

\begin{figure}
    \centering
    \includegraphics[width=0.48\textwidth]{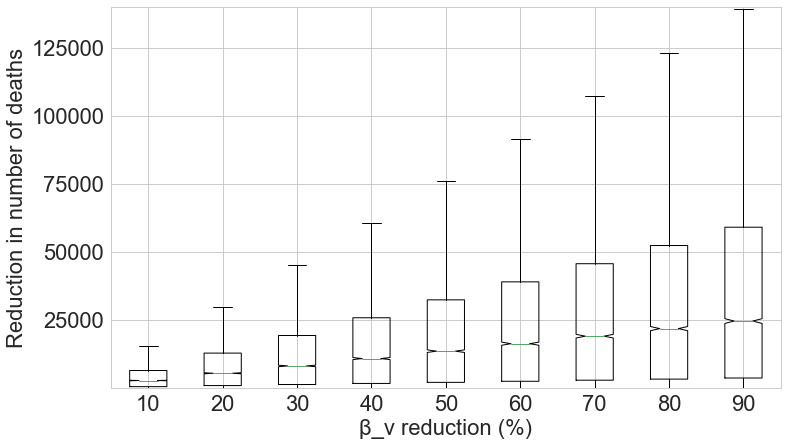}
    \caption{Reduction in the number of deaths as a function of the percentage decrease in $\beta_v$, given that this reduction is applied from 2 June 2020.}
    \label{fig:betaredwholenow}
\end{figure}

The potential reduction in the number of deaths caused by COVID-19 as a function of the percentage decrease applied to $\beta_v$ achieved by the Monte Carlo simulation is shown in Figure \ref{fig:betaredwholenow}. As it can be seen in the figure, the reduction of the exposure of vulnerable groups to the disease from today (2 June 2020) greatly decreases the number of further deaths, The mean figures for the prevented deaths for each 10\% decrease in $\beta_v$ are shown in Table \ref{tab:deathdecreasenow}. 

\begin{figure}
    \centering
    \includegraphics[width=0.48\textwidth]{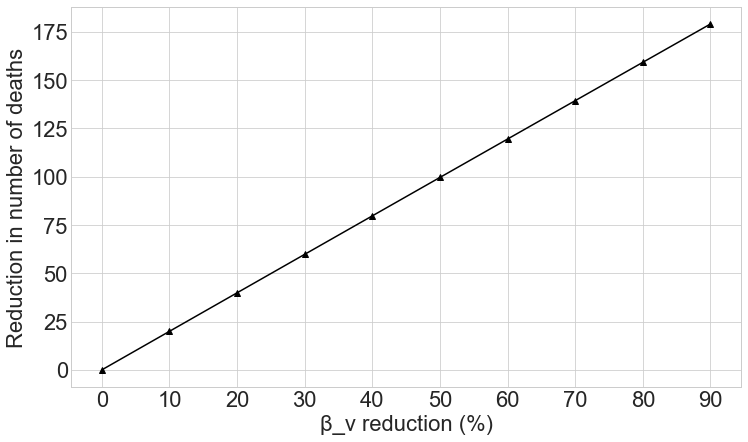}
    \caption{Reduction in the number of deaths as a function of the percentage decrease in $\beta_v$ for the best fit model, given that this reduction is applied from 2 June 2020.}
    \label{fig:rednowbestfit}
\end{figure}

The results obtained by the use of the best fit model are shown in Figure \ref{fig:rednowbestfit}.

\begin{table*}[]
\caption{Relationship between the decrease in $\beta_v$ and the resultant number of deaths avoided when this decrease is applied from today (2 June 2020) expressed as the mean value of the Monte Carlo simulation.}
\centering
\renewcommand{\arraystretch}{1.8}
\begin{tabular}{|l|l|l|l|l|l|l|l|l|l|}
\hline
Decrease in   $\beta_v$           & 10\% & 20\%  & 30\%  & 40\%  & 50\%  & 60\%  & 70\%  & 80\%  & 90\%  \\ \hline
Decrease in number of deaths & 3681 & 7406 & 11172 & 14975 & 18810 & 22673 & 26559 & 30464 & 34383 \\ \hline
\end{tabular}
\label{tab:deathdecreasenow}
\end{table*}

\section{Discussion} \label{sec:recommendations}
From the results of the modelling, it has been evidenced it is crucial to consider the contrasting vulnerability in different groups of the population regarding their age and overall health condition when developing plans to return to normality. Motivated by this, three main points for consideration are proposed, namely 1) the use of dedicated wristbands for the vulnerable to enable social distancing and protect their well-being, 2) the use of Personal Protective Equipment(PPE) and 3) the lockdown easing plan and the benefits they can bring. 

\subsection{Protecting the Vulnerable People}
Contact tracing, which has been used alongside other protection measures across the world, can keep a record of any new infection cases and anyone who has been close to them \cite{WHOtracing}. This could enable uninfected and immune people to leave their homes, while people who might have been infected instructed to self-isolate. However, contract tracing is time consuming and resource intensive and to be strictly accurate and valid requires 100\% adoption.
Bluetooth technology has been used previously for messaging and tracking of nearby devices using proximity detection \cite{Viralnet} \cite{kanjo12}. Apple and Google along with many health authorities have proposed software smartphone hosted apps using Bluetooth Low Energy (BLE) to automate the contact tracing process \cite{Apple}. BLE is a form of wireless communication designed especially for short-range communication suitable for situations where battery life is preferred over high data transfer speeds. 
Unfortunately, vulnerable and older people are more likely to use older smartphones that don’t come equipped with the BLE feature, e.g. about 9-12\% of smartphones in the UK lack the BLE functionality needed for it to work.
In addition, data from Ofcom shows that while around 80 per cent of all adults owned a smart mobile phone in 2018, only 47\% of 65-74 year old's, and 26\% of over 75s did \cite{Ofcom}. This is supported by data published by Statista, which indicates that only 40\% of over 65s used a smart phone to connect to the internet in 2019 \cite{statista}. Furthermore, many people have privacy concerns about using their smart phones as a tracer and they might not be willing to download a contact tracing app.
With this background, three possible solutions to support vulnerable people are proposed 1) contact Tracing wristband or wearable,2) social distancing alert mechanism and, 3) a wearable to monitor symptoms.

\subsubsection{Wearables for Contact Tracing}
The effectiveness of contact tracing hinges on how many people use it. It is proposed that governments could provide vulnerable individuals with a BLE wristband similar to the one in Figure \ref{fig:wristband}), closing a data collection hole created by systems relying on smartphones and an app only. Aiming to take a population to 100

\begin{figure}[t]
  \centering
  \includegraphics[width=0.48\textwidth]{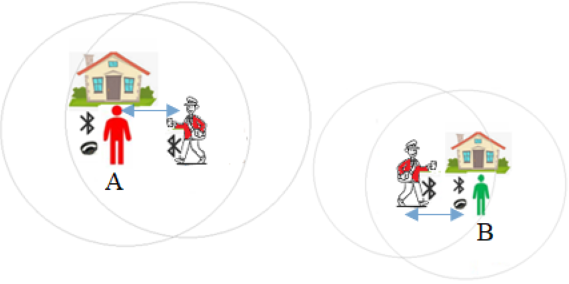}
  \caption{A schematic of wearable-based COVID-19 proximity tracing}
  \label{fig:tracing}
\end{figure}

The wristband includes a proximity sensor powered by BLE. It also includes a manual control to self-report and change a wearer’s status, recording states like self-isolating, symptomatic and, tested negative or tested infected. When a user updates their state to indicate an infection after testing, that updates others they have been in close proximity with. 
\begin{figure}[H]
  \centering
  \includegraphics[width=0.48\textwidth,height=0.15\textheight]{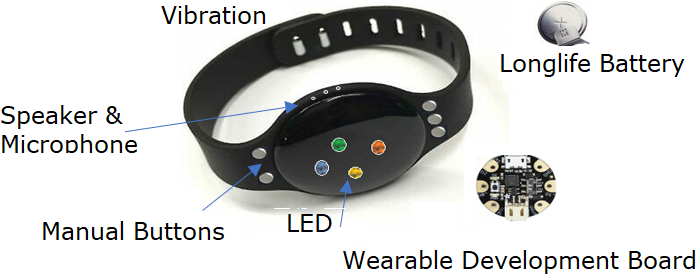}
  \caption{Wristband for Vulnerable People}
  \label{fig:wristband}
\end{figure}
Figure \ref{fig:tracing} shows how the wristband concept works. Contacts of an individual wearing the wristband A are recorded on the wristband. A passer-by (e.g. postman) with a smartphone and Contact Tracing app comes within read range (longer range than close proximity) and downloads the records broadcasted by the wristband A. In turns. The individual wearing the wristband A could be alerted in real time using LED light (or sound) of close proximity events.
We believe that adopting wearable (e.g. wristbands) solutions is advantageous over mobile apps for the following reasons:
\begin{enumerate}
    \item Mobile phones might not be always with users. Instead, they might be left at home, in the car or at work, which means their social encounters don't always correspond to actual contact.
    \item A wristband solution will only have radio technology with a small memory and a battery with no access to users data which could help to preserve privacy, as it is low power it can always be on. It does not require setting up or installation by the wearer.
    \item Phone apps require to be installed and activated by the users, Bluetooth need to be switched on. These requirements make the apps unfeasible for vulnerable people with difficulties in remembering instructions; and lack of digital literacy, vision or motor control.
    \item Contact Tracing app can consume more energy as they are often kept active with battery optimisation features disabled.
    \item Wearables are more likely to be work in front of the body (e.g. wristbands, necklace or a keyfob), which could potentially improve the accuracy of the proximity detection in the case of the face of face contact.
    \item Smart phones come with different operating systems and settings, which means each model might require individual calibration and configuration.
\end{enumerate}

\subsubsection{Digital tools to maintain Social distancing }
By combining very dense contact tracing data from smartphone apps and wristbands signals with information about infection status and symptoms, vulnerable people can be protected and kept safe.
Many countries have introduced wristbands for different purposes. For instance, in South Korea, people found to be violating lockdown rules can be ordered to wear a tracking band, which alerts the police if people leave the house \cite{bbc_2020}. The trackers were introduced after people started to leave their phones at home to avoid detection. The devices also alert the authorities if people try to remove it. Bulgaria has been testing Comarch LifeWristbands, developed in Poland \cite{bbc_2020}. This system in addition to confirming a person is staying at home, can monitor the wearer's heart rate and then be used to call the emergency services.

\begin{figure}[t]
  \centering
  \includegraphics[width=0.48\textwidth]{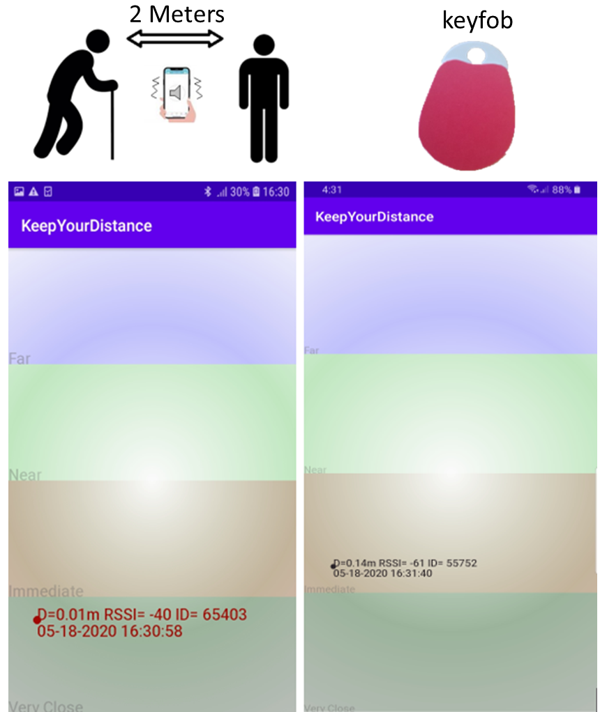}
  \caption{A BLE mobile app and a keyfob concept developed at Nottingham Trent University (NTU) to alert people when they are within 2m proximity.}
  \label{fig:socialdistancing}
\end{figure}

Contact Tracing apps and Internet of Things (IOT) such as key fobs, tags or wrist bands can also be used to alert people (e.g. using vibration) if another device comes within a specified distance. Figure \ref{fig:socialdistancing} shows BLE keyfob and KeepyourDistance app concepts being developed at Nottingham Trent University (NTU) to maintain social distancing.
The KeepyourDistance screenshots show the signal strengths of nearby devices and it vibrates when other phones with the same app are in close proximity.

\subsubsection{Wearable to monitor symptoms}
Dedicated health wristbands can be provided to vulnerable people with underling health conditions to track their health including: temperature, breathing and heart rate, and transmit it to their doctors.
In addition to the above functionalities, most commercially available smart-wristbands and smart-watches incorporate an inertial measurement unit (IMU) composed of tri-axial accelerometers and gyroscopes. The work in \cite{Kwok2015} reports a face touching frequency of 23 times per hour with 44\% of the face touches involving contact with a mucous membrane. Wristbands should be programmed to incorporate an alert mechanism to warn users whenever a potential movement of the hand towards the face is detected. The accuracy achieved by relevant gesture recognition work using inertial sensors \cite{Anderez2019,Anderez2020} suggests that the development of such alerting features is feasible.   
These wearables can then be re-purposed once the COVID-19 epidemic is over. For example they can be given to older people and individuals with underling heath conditions to monitor their health and well-being. Also they can be used to keep vulnerable people connected with local volunteers and community services.

\subsection{Disease Transmission and the Use of Personal Protective Equipment}
As outlined by the WHO \cite{World2020transmission}, COVID-19 is primarily transmitted from person to person through small droplets exhaled by the mouth or nose when an infected individual speaks, sneezes or coughs. Direct transmission then occurs when such droplets travel onto the mucous membranes of susceptible recipients, necessitating contact at close range (usually within 1 meter) \cite{Leder2005}. In addition to this, exhaled droplets can also come to rest on surfaces around the infected individual. According to the WHO, the COVID-19 virus can survive up to 4 hours on copper, 24 hours on cardboard and up to 72 hours on plastic and stainless steel surfaces. As a consequence, these become a potential source of infection when touched by susceptible individuals prior to touching their nose, mouth or eyes. 
Although exhaled droplets are often heavy enough not to travel great distances, sinking quickly to the ground or surrounding surfaces, viral particles in the form of bio-aerosols can remain airborne for an extended period of time, particularly when droplet diameters are either too small for gravitational deposition ($<$2 $\mu$m) or too large for diffusive deposition ($>$200 $\mu$m) \cite{Stahlhofen1989}. Experimental results on SARS during the 2003 epidemic \cite{Booth2005} support these statements, showing viral particles in the form of bio-aerosols were being emitted by hospitalised patients. Likewise, recent work by \cite{Van2020} demonstrates aerosol COVID-19 remains viable and infectious with a half-life on the order of 1 hour, thus confirming the plausibility of its transmission via airborne particles.

Given the above, the use of personal protective equipment (PPE) can be a key aspect to prevent the transmission of COVID-19. In this context, PPE includes, among other equipment, the use of respiratory protection to safeguard the mucous membranes and to prevent the transmission of viral droplets such as masks and respirators, as well as that of physical barriers such as gloves, goggles and face shields.

\subsubsection{The Use of Masks and Respirators}\label{subsec:masks}

According to the European standards EN149:2001 and EN14683, there are four types of filtering masks, namely Filtering Face Piece 1 (FFP1), FFP2 \footnote{The American and Chinese equivalents for FFP2 are N95 and KN95 respectively}, FFP3 and surgical masks, with different models differing primarily in the filtration efficiency given by their capability to filtrate inwards and outwards particles. For the scope of this work, the use of surgical and FFP2 masks is discussed. In terms of inwards filtration, surgical masks are designed to protect against droplets, sprays and any other particle with a diameter greater than 100$\mu$m \cite{Smereka2020}. FFP2 masks, however, retain $>$94\% of the particles smaller than 0.5$\mu$m. As the work in \cite{Bake2019} indicated, exhaled particles can range from 0.01 to 1000 $\mu$m, with COVID-19 particles exhibiting a round or elliptic shape with diameters ranging from 0.06 to 0.14 $\mu$m \cite{Cascella2020}. As a consequence, surgical masks can prevent the inhalation of COVID-19 particles when these are expelled in the form of droplets with diameters greater than 100$\mu$m but not when expelled in the form of small airborne particles. In contrast, FFP2 masks or respirators prevent the inhalation of both droplet and airborne viral particles. Despite the drawbacks encountered on surgical masks when it comes to the prevention of the filtration of small airborne particles, it must be noted that they can reduce the emission of viral particles into the environment \cite{Leung2020}. It is worth noting that mask effectiveness decreases with increasing concentrations of water vapour and carbon dioxide between the face and the mask/respirator caused by each subsequent exhalation \cite{Smereka2020}. Thus, masks should be replaced frequently. I should also be noted face masks have to fitted correctly and form a seal peripherally to stop air passing around the mask and not through it.

It should also be stated, the incorrect use of PPE, such as not changing disposable masks or gloves, can have a counterproductive effect, thus jeopardising their protective effect and even increasing the risk of infection \cite{Feng2020}. Given this, health organisations and government bodies should be spreading good clear information covering how to wear and discard the different recommended protective equipment components properly. 

\subsubsection{Other Personal Protective Equipment}\label{subsec:otherPPE}
As mentioned above, although COVID-19 is mainly transmitted through direct contact between an infected and a susceptible individual, indirect transmission is also plausible. In addition to masks and respirators, the use of further PPE can help to reduce the risk of the transmission of COVID-19 both directly and indirectly. In this context, face shields and eye protection equipment such as goggles can play an important role in preventing both direct and indirect transmission. 

A face shield is a PPE component which provides a physical protective barrier to the facial area and related mucous membranes \cite{Roberge2016}. Various experimental works \cite{Lindsley2014,Shoham} have shown the potential effectiveness of face shields against the transmission of viral respiratory diseases like COVID-19. For instance, the work in \cite{Lindsley2014} employed a cough aerosol simulator filled up with influenza virus (aerosol mean diameter of 8.5$\mu$m) alongside a breathing simulator to test the effectiveness of a face shield against the transmission of the virus. The results reported outline risk reductions of 96\% and 92\% on the inhalational exposure right after a cough at distances of 46cm and 183cm respectively. Reducing the aerosol diameter to 3.4$\mu$m resulted in a reduction of the blocking effectiveness to 68\% at 46 cm right after a cough and to a 23\% over a post-cough period of 1 to 30 minutes. The experimental work in \cite{Shoham} used a fluorescent dye
(particle diameter $\approx 5\mu m$, distance $\approx 50cm$) to evaluate the effectiveness of a face shield at protecting eye contamination from aerosol particles. The results outlined the use of a full face shield (FFS) completely prevented eye contamination. Additionally, it was demonstrated that the combination of the FFS with a N95 respirator offered full protection of the eyes, nose and mouth from contamination. In contrast, the use of safety glasses with either a surgical mask or N95 respirator resulted in some eye contamination. 

According to WHO recommendations \cite{World2020PPE}, gloves should be worn by medical personnel, laboratory technicians manipulating respiratory samples, caregivers who provide direct care to confirmed COVID-19 patients and visitors entering a COVID-19 patient's room. Therefore, the WHO does not consider the general use of disposable gloves by the community necessary. Instead, a thorough hand hygiene is proposed. Besides hand hygiene, a consistent disinfection of commonly touched surfaces is recommended when an individual is quarantined after contracting the disease to prevent its spread to susceptible individuals who may co-handle those surfaces. In line with the WHO, it should not be recommended that the general public use disposable gloves, as this is unlikely to reduce the risk of infection and it may lead to a potential unnecessary shortage for scenarios in which they are indispensable and absolutely necessary.  

\subsubsection{Final Recommendations on the Use of PPE}
Provided the analysis upon the effectiveness of the different PPE components on the prevention of the spread and on the contraction of COVID-19 carried out in Sections \ref{subsec:masks} \ref{subsec:otherPPE}, the following recommendations are made:
\begin{enumerate}
    \item Masks should be worn by every individual in the population. Non-vulnerable individuals should use surgical masks. FFP2 masks or respirators should be worn by vulnerable individuals, especially in closed environment public spaces such as supermarkets, pharmacies and public transport.
    \item Face shields are recommended for vulnerable individuals, especially in closed spaces. Ideally, face shields should be worn in conjunction with an FFP2 mask or respirator.
    \item Gloves are not recommended for the general public.
\end{enumerate}

\subsection {Adoption and Incentive Mechanisms for Behavioral Change}
\begin{figure}[t]
  \centering
  \includegraphics[width=0.48\textwidth]{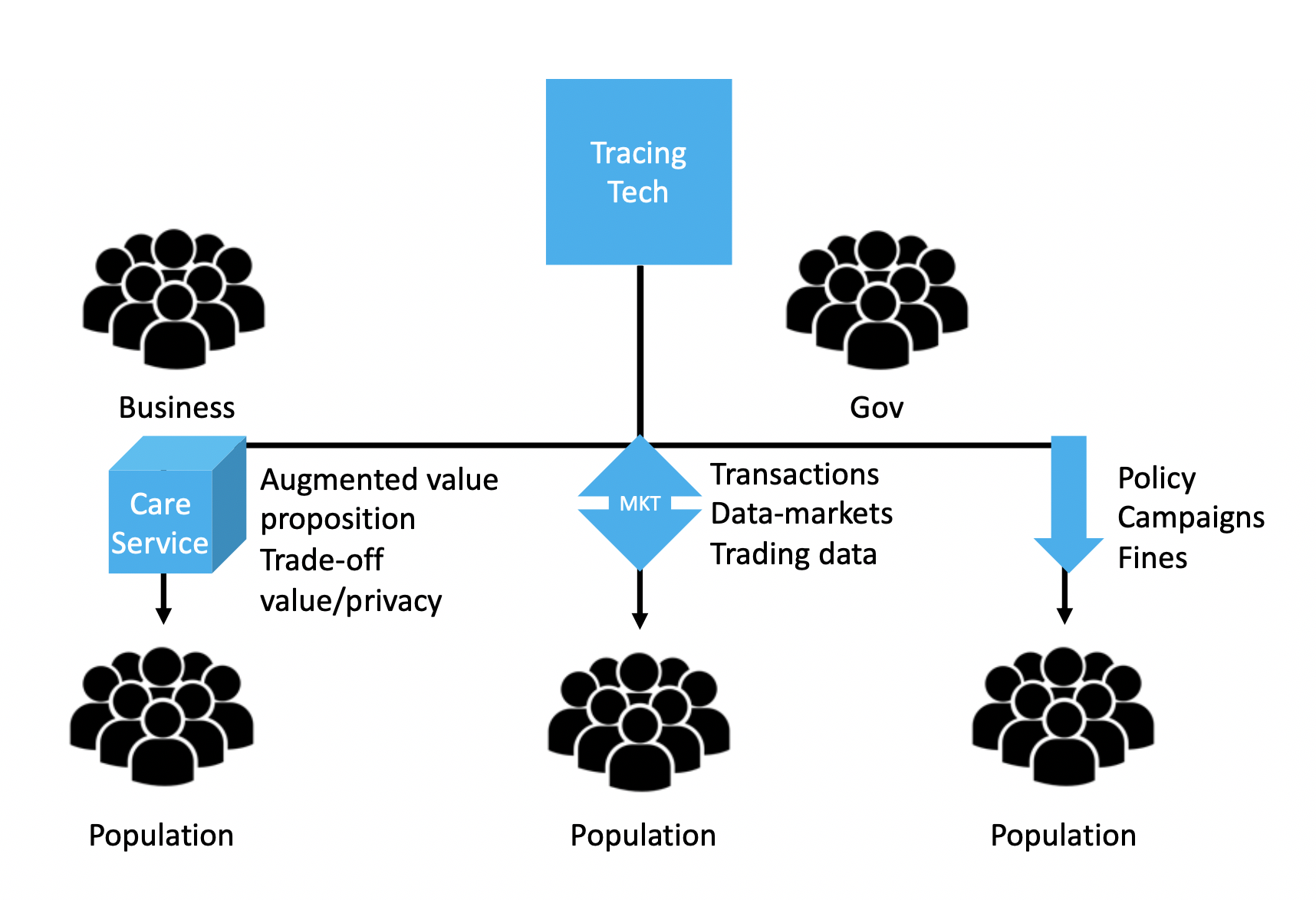}
  \caption{Adoption Approaches}
  \label{fig:adoption}
\end{figure}

One of the challenges around implementing the proposed solutions is getting people to adopt and use them. In analogous settings, when young drivers were encouraged to use tracing technology solutions to improve their driving habits; young drivers, who participated in the technology trials, only used behaviour monitoring apps while the incentives lasted and stopped when the incentives stopped. Moreover, drivers often tried to play and work the system in order to obtain extra incentive points \cite{musicant2016can}. While the need to motivate individual adoption is universal, a segmented reaction to the proposed solutions and best practise is anticipated, so different strategies and incentive mechanisms might be required for different groups. Successful adoption depends on individual motivation, incentives provided through a variety of strategies. The incentive mechanisms can range from providing information that is meant to resonate with basic values, such as “using this technology will save lives” to material and non-material incentives in the form of (i) paying people for their personal data (e.g., providing tax-rebates, etc.) or (ii) providing priority access to some services (e.g., medical services, etc.). For example, in the past, people successfully adopted healthy behaviours when they received financial incentives \cite{marteau2009using}. From this, it follows that if incentives are offered people (a) adopt a certain behavior or (b) engage in a behavioral change. Rightly or wrongly these incentives may be more effective if they appeal to people's "present bias", which is the tendency to pursue smaller and more immediate rewards (e.g. getting a small amount of money every day) rather than bigger and more abstract goals, such "eradicating COVID-19 in the world".
One potential incentivization mechanism could be offering a better value proposition to the user that goes beyond tracing. Considering that the vulnerable population generally tend to procure care services more often \cite{shivayogi2013vulnerable}, the function of contact tracing is likely to succeed if it is embedded into the general care applications and services. An example of such general care services would be framing the contact tracing technology as a “digital nurse” for the vulnerable groups, that aims to monitor (in real time) the state of the wearer and potentially offer some desirable care features. Features such as health monitoring or being able to make automated calls for medical help in case the wearable detects signs of distress, etc. \cite{vegesna2017remote}. Having a clear value proposition, which would go beyond the functional purpose of tracing would allow the technological solutions to succeed with the elderly population \cite{roupa2010use} more easily, as users will see not only how their data can benefit society, but also how their data can help them receive better, more efficient, and higher quality care. In such situations, due to the behavioural "privacy paradox" \cite{norberg2007privacy}, users, who are concerned about privacy, are also likely to perceive the added value of the technological solutions as a trade-off between their personal data and desirable services. If the benefits of using the technology outweighs the cost (in terms of personal data loss), users will be much more likely to adopt the new wearables, even if that requires disclosing their location \cite{xu2011personalization}. This, in turn, will significantly increase the chances of technology adoption as well as ensure policy success. Considering that people tend to attribute greater value and use products that have a cost associated with them (e.g., people see greater value in goods and services they need to pay for rather than free goods and services), such value proposition (i.e., offering medical or care service benefits in exchange for personal data) is likely to be self-sustainable.
An alternative incentive mechanism can be implemented through material incentives by offering to buy personal data from the users (i.e., the users will receive financial remuneration for their personal data) \cite{schwartz2003property}. This mechanism will leverage on participants' willingness to trade privacy in the context of willingness to accept (WTA) money in exchange for data. The mechanism could work through offering a certain amount of money or an equivalent in discounts for users' personal data in a clear and transparent transaction \cite{bizon2016willingness}. The key to success in this case is full transparency about (i) who is the data buyer (government or third-party organizations); (ii) how the data buyer will use the data; and (iii) how the data buyer will protect user data.
The third strategy to push for adherence to the use of the proposed solutions, is through legal means. For example, contact tracing could be mandated by law and distributed to the eligible population. The use of the technology can then be monitored and "fines" for non-compliance could be introduced. The "fines" do not necessarily have to be monetary. They can be implemented as decreased benefits. The success of such this approach greatly depends on cultural values and social norms (in some societies punishment may work better than reward in policy implementations) \cite{milberg1995values}.
The three suggested approaches are summarized on Figure \ref{fig:adoption}.

\subsection{Lockdown Easing Plan}
Once past the peak of the epidemic, that is, the exponential phase of the curve, it is then time to develop plans for a gradual easing of the imposed restrictions. 
Several research works \cite{Kermack1927,Aron1984,Ferretti2020}, including the work presented here, have proposed a variety of epidemiological models to study the spread of the disease and to make the consequent estimations of the total number COVID-19 cases in different territories. Having done that, it is possible to have estimations on the herd immunity gained throughout the progression of the disease. However, there are still many unknowns of the disease which may potentially compromise the estimations made by the application of such epidemiological models. Given this, government bodies should be pressed to carry out prevalence studies to further support the estimations obtained by the use of epidemiological models, as it is crucial for the minimisation of further infection cases and consequently for the reduction of the number of further deaths caused by the disease, to have accurate numbers for the number of people who have already had and recovered from the disease and have, thus produced antibodies against it.   

\subsection{Conclusions and Future Work}\label{sec:conclusions}
The key driver for this work has been the large size of the vulnerable population and their higher risk of severe infection and death. This group is also prone to suffer from loneliness resulting from the prolonged period of lockdown.
Governments need to balance the need to socially isolate vulnerable people and shield them while also taking into account their mental health and well-being which might be severely affected by the isolation measures.

A modified SEIR model, namely SEIR-v, through which the population was separated into two groups regarding their vulnerability to the disease was proposed to provide a means of studying the spread and the case fatality rate of COVID-19 when different contention measures are applied to different groups regarding that vulnerability. Using SEIR-v, the impact of a reduction in the exposure of vulnerable individuals to COVID-19 on the number of fatalities caused by the disease was analysed. The results indicate an average of 3681 deaths can still be saved by only reducing by 10\% the exposure of vulnerable groups to the disease from now. In line with this and also considering the negative consequences caused by the application of strict isolation measures on people's mental health, a set of recommendations including the adoption of digital tools and protective equipment was proposed. 
Future work will be directed towards the analysis of the lockdown easing steps taken by the U.K. government and their potential impact on further fatalities caused by COVID-19.


%
\section*{Conflict of interest}
%
On behalf of all authors, the corresponding author states that there is no conflict of interest.

\bibliographystyle{spmpsci}      
\bibliography{bibliography.bib}   

\end{document}